# FINITE ELEMENT SIMULATION OF TEXTILE MATERIALS AT THE FIBER SCALE

Damien Durville
CNRS – UMR8578 / Ecole Centrale Paris
Grande Voie des Vignes
92295 Châtenay-Malabry – France
damien.durville@ecp.fr

## ABSTRACT

A general approach to simulate the mechanical behavior of textile materials by taking into account all their constitutive elementary fibers and contacts between them is presented in this paper. A finite element code, based on an implicit solver, is develop to model samples of woven fabrics as assemblies of beams submitted to large displacements and developing contact-friction interactions between themselves. Special attention is paid to the detection and modeling of the numerous contacts occurring in this kind of structures. Robust models and efficient algorithms allow to consider samples made of few hundreds of fibers. The unknown initial configuration of the woven fabric is obtained by simulation according to the chosen weaving pattern, providing useful informations on fibers trajectories and tows cross-sections shapes. Various loading cases can then be applied to the textile sample, in presence of an elastic matrix, so as to identify its mechanical behavior under different solicitations. Results of simulations performed on coated fabric samples are given to illustrate the approach.

## 1. INTRODUCTION

Because the global behavior of textile composites depends on local phenomena occurring within the fabric at the scale of tows or fibers, a good understanding of mechanisms taking place at the scale of fibers is important. The nonlinear behavior of dry fabric or fabric-reinforced composites often originates in phenomena such as transverse compression of yarns or tows, sliding between tows or fibers, or locking between fibers. Predicting these phenomena is very hard without focusing the analysis at the scale of fibers.

A textile composite is made of components of different level. At a mesoscopic level, it can be seen as an assembly of yarns or tows, coupled with an elastic matrix. Modeling the structure at this mesoscopic level requires to be able identify the global behavior of tows or yarns. Because it involves nonlinear couplings between loads and strains in different directions due to interactions between fibers, the behavior of yarns is complex to identify. Going down to the scale of fibers allows to bypass the identification of intermediate models at upper levels and to base the global model only on mechanical properties of fibers and matrix and of contact-friction interactions between fibers.

Recent improvements in both computing capacity and modeling and algorithmic aspects, make the finite element method an appropriate tool to study textile composites. This paper presents a general approach, based on this method, to model the behavior of samples of dry fabric or textile composite materials, taking into account all elementary fibers together with the interactions developed between them and the coupling with an elastic matrix.

As an additional difficulty, when dealing with woven structures, the geometry of the initial configuration of the fabric (individual fiber trajectories) can not be known explicitly. Before starting with applying various loading cases on textile samples, the first task of the simulation is therefore to compute this initial geometry according with the weaving pattern.

The global problem is set as to find out an equilibrium solution for an assembly of fibers submitted to large displacements, developing contact-friction interactions, and bound to an elastic matrix. Fibers are represented by a 3D beam model, formulated in a large displacements and finite strains framework. The quasi-static incremental problem is solved using an implicit solver, which allows to consider large loading increments. The use of an implicit solver with many nonlinearities requires iterations until finding a solution satisfying the equilibrium and may lead to a high computation time. Because of the huge numbers of fibers and contacts, efficient and robust algorithms are needed to reduce this computation time.

Some similar approaches can be found in the literature. At the scale of fibers, for the computation of the initial configuration for braided structures or 3D interlock woven fabric, digital elements have been used by Miao and al. [1]. Since these digital elements have neither bending nor torsional stiffness, fibers must be tightened to find a solution. Finckh [2] proposed to simulate the weaving process and to apply dynamic loading cases using an explicit solver. Other approaches tackle the problem at the scale of yarns, representing yarns by beams or 3D models and studying interactions between them [3,4,5]. The simulation approach developed in this paper has been previously applied to other kinds of entangled media [6,7].

To introduce the global approach, a first section is dedicated to the beam model employed to represent fibers. The contact modeling, including geometrical and mechanical aspects, is then described. In a third section, the way the initial configuration is computed according to the weaving pattern is explained. After presenting how an automatically meshed elastic matrix can be added and coupled to the woven fabric, results of different loading tests on dry fabrics and composites, for both plain weave and twill weave are given.

## 2. 3D BEAM MODEL

2.1 Introduction

Beam models are usually based on the assumption that their cross-sections remain rigid. The motions of these cross-section can therefore be described by two kinematical fields and six degrees of freedom : one translation of the center of the section, and one rotation. However, these usual models have two main disadvantages : the handling of large rotations requires complex formulations, and they can not account for deformation of cross-sections. To overcome these difficulties, we use a richer model described by nine degrees of freedom.

## 2.2 Enriched kinematical beam model

The enriched kinematical beam model is based on a Taylor expansion of the placement of particles of the beam with respect to the line of centroids. If we denote $\mathbf{z}\,(z_1,z_2,z_3,)$ a material particle defined in the reference configuration (see Fig. 1), the placement $\mathbf{x}\,(\mathbf{z})$ of this particle can be expressed as follows :

$$\mathbf{x}\,(z_1,z_2,z_3) = \mathbf{x}\,(0,0,z_3) + z_1 \partial_1 \mathbf{x}\,(0,0,z_3) + z_2 \partial_2 \mathbf{x}\,(0,0,z_3) + \mathbf{o}(z_1,z_2). \qquad [1]$$

Denoting respectively $\mathbf{x}_0$, $\mathbf{g}_1$ and $\mathbf{g}_2$ the three kinematical vectors introduced as first terms of the above expansion, we assume as kinematical model that the placement of any material particle of the beam is expressed as function of these vectors in the following way :

$$\mathbf{x}\,(z_1,z_2,z_3) = \mathbf{x}_0\,(z_3) + z_1\,\mathbf{g}_1\,(z_3) + z_2\,\mathbf{g}_2\,(z_3). \qquad [2]$$

With this model, three vectors are used to describe any cross-section of the beam : one vector to give the position of its center, $\mathbf{x}_0$, and two other vectors, $\mathbf{g}_1$ and $\mathbf{g}_2$, called section vectors, which determine the orientation of the section. Corresponding to this model for the placement, we assume that the displacement $\mathbf{u}\,(\mathbf{z})$ of any particle is also expressed by means of three vectors as follows :

$$\mathbf{u}\,(z_1,z_2,z_3) = \mathbf{u}_0\,(z_3) + z_1\,\mathbf{h}_1\mathbf{x}\,(z_3) + z_2\,\mathbf{h}_2\,(z_3). \qquad [3]$$

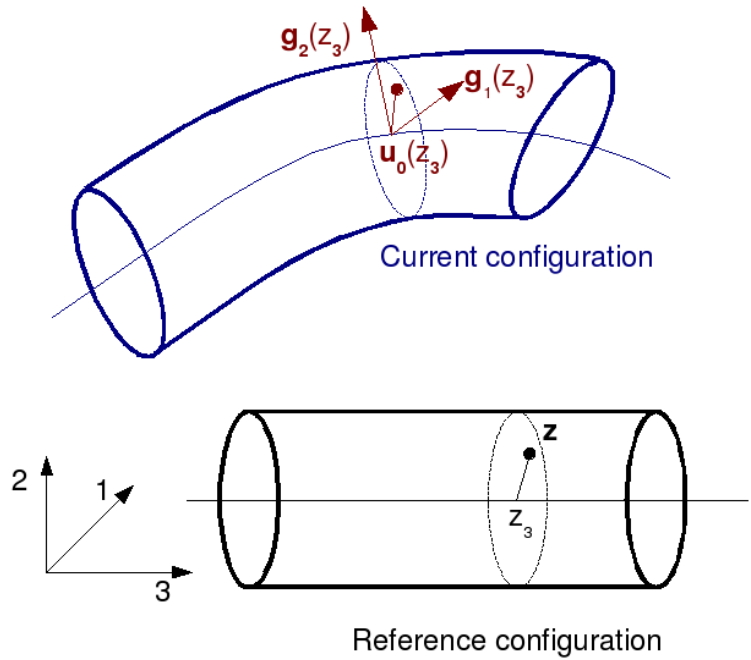

Figure 1. The three kinematical vectors used to define the placement in the kinematical beam model.

Since no conditions are set for the two section vectors, neither in norm, nor in angle, plane deformations of cross-section can be considered by the beam model, in addition to usual deformations (stretching, torsion, bending and shear).

2.3  Strains and constitutive law for the beam model

With the above kinematical beam model, because it is able to reproduce plane deformations of cross-section, no term of the Green-Lagrange strain tensor is constrained to be zero, and fully 3D effects, such as Poisson's effect (strains in transverse directions induced by stretching), can be caught. Because the strain tensor is fully three-dimensional, an usual 3D elastic constitutive law, taking into account strains in all directions, can be used to model the behavior of each beam.

## 3.  CONTACT MODELING

3.1  Introduction

Contact modeling is crucial point since contact-friction interactions make the specificity of the behavior of textile structures. The main difficulties are related to the fact that contact areas can be numerous, and are constantly evolving, as contacts may appear or disappear anywhere and at any time.

3.2  Geometrical detection of contact

Since we are working in a large displacements framework, and because the rearrangement of fibers during the simulation, the process of contact detection needs to be regularly repeated during the calculation. For this reason, this process must be efficient and with low CPU cost.

Contact configurations between fibers can be very different depending on their relative positions. Very localized and reduced contact areas can be observed at crossings between fibers from two different tows, but there can also be long and continuous contact lines between two neighboring fibers of the same tow.

To face these different configuration, we suggest a general method to handle contact, based on the generation of contact elements, made of pairs of material particles attached to the surface of fibers, and based on geometrical criteria.

Whereas in usual methods, contact interactions are determined by fixing a first point on one of the structures and by searching a target point on the opposite geometry, using generally a normal direction to one surface, we suggest do define contact elements from an intermediate geometry. This geometry has to be generated for each region where two parts of fibers are close enough and where contact is likely to produce. This way of searching contact from a third geometry offers a symmetrical treatment of contacting structures.

The first step to create contact elements is to determine in the global collection of beams, a set of proximity zones (see Fig. 2), defined as pairs of parts of lines holding beam meshes, which are close to each other. The determination of proximity zones is performed, for each couple of beams in the assembly, by distributing test points on one of the lines and calculating the nearest point on the opposite line.

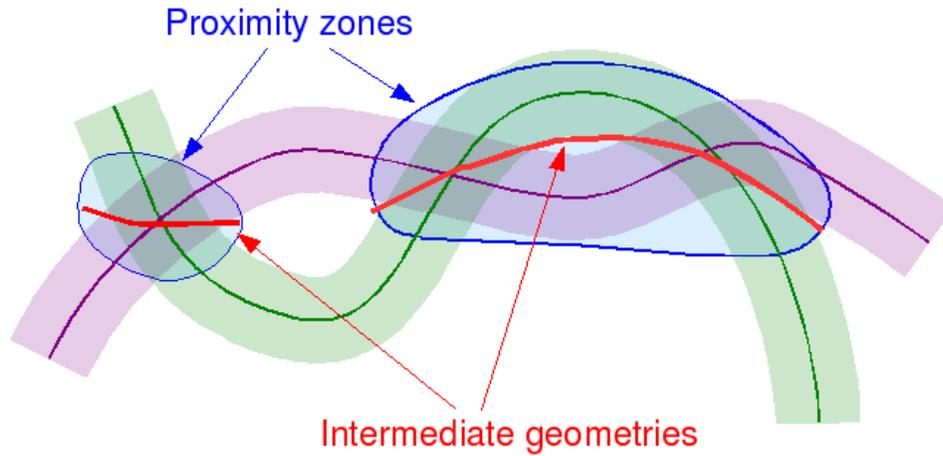

Figure 2. Definition of intermediate geometries based on proximity zones.

The intermediate geometry is then defined, for each proximity zone, as the average between the two close parts of lines. Contact elements are generated from this intermediate geometry, using planes orthogonal to this geometry to determine couples of beam cross-sections candidates to contact. Material particles are then accurately located on the border of these cross-sections.

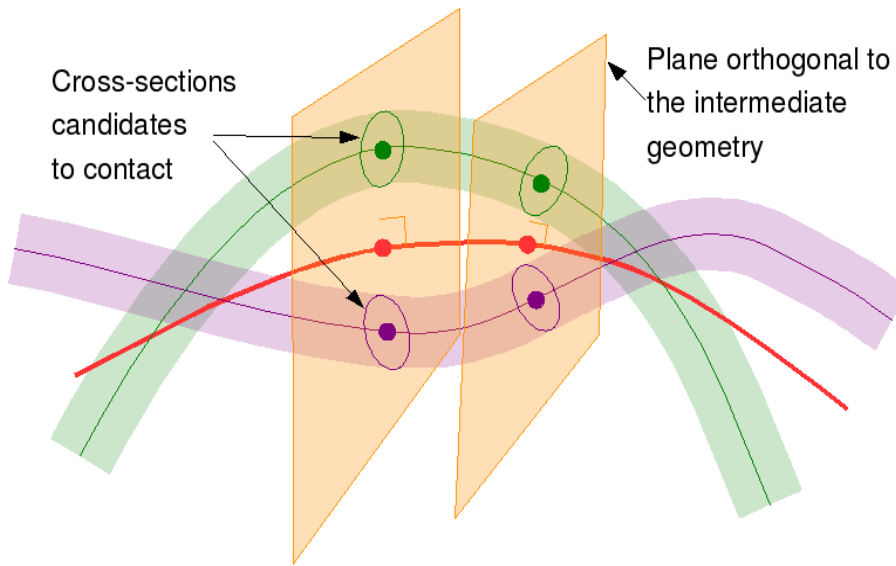

Figure 3. Determination of couples of cross-sections candidates to contact.

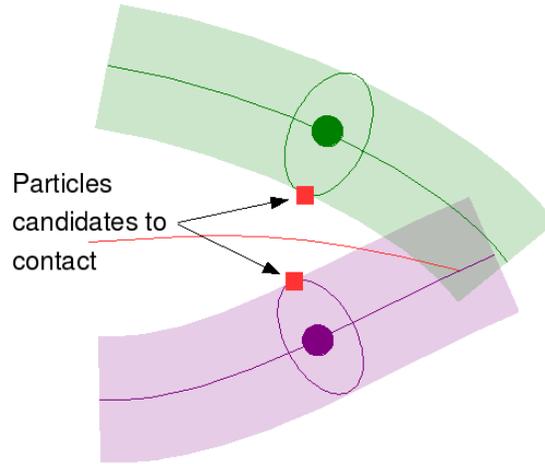

Figure 4. Selection of material particles on the borders of cross-sections to create a contact element.

### 3.3 Definition and generation of contact elements

We define a contact element $E_c^{(ij)}(\mathbf{x}_G)$ as a pair of material particles, $\mathbf{z}^{(i)}$ and $\mathbf{z}^{(j)}$, belonging to two different beams (i) and (j), and predicted to enter into contact at a given location of the intermediate geometry :

$$E_c^{(ij)}(\mathbf{x}_G) = (\mathbf{z}^{(i)},\mathbf{z}^{(j)}) \text{ such that : } \mathbf{x}(\mathbf{z}^{(i)}) = \mathbf{x}(\mathbf{z}^{(i)}) = \mathbf{x}_G. \qquad [4]$$

The equality between positions in the above equation must be understood in the sense of a prediction.

The discretization of the contact problem is performed by distributing on all intermediate geometries in the structure series of discrete locations where contact element are created. The fact that the construction of contact elements depends on the geometry of fibers, and therefore on the displacement solution of the problem, introduces in the formulation an additional nonlinearity, and requires the determination of contact elements to be repeated as the solution evolves.

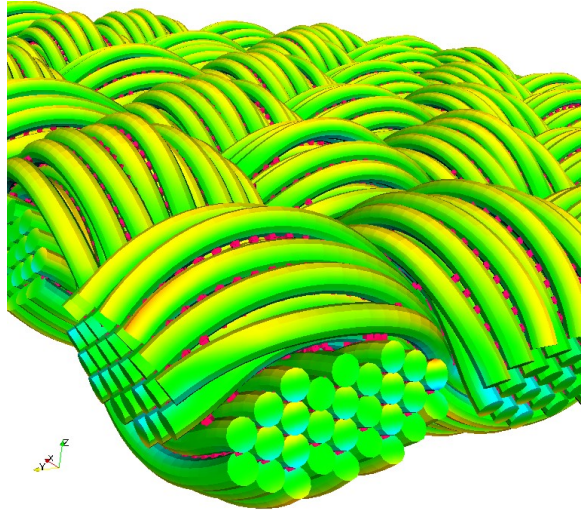

Figure 5. Contact elements (red spots) generated in a woven structure

3.4 Formulation of contact conditions

### 3.4.1 Adding of a normal contact direction

In order to formulate linearized contact conditions, a normal direction, denoted $\mathbf{N}\ (E_c^{(ij)})$, is associated to each contact element $E_c^{(ij)}$. This direction sets the direction according to which distance between interacting particles will be measured.

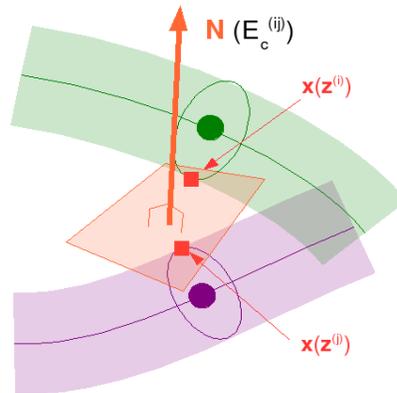

Figure 6. Definition of a normal direction for a contact element.

This contact direction is generally taken as the one between the two centers of cross-sections involved in contact. Once again, such a definition introduces a nonlinearity in the global formulation since the normal contact direction depends on the solution itself.

### 3.4.2 Kinematical contact condition

In order to prevent interpenetration between beams, kinematical conditions are set for all contact elements. They aim at imposing that the distance between particles of a contact element, measured according to the normal contact direction, remains positive. This distance, defined as the gap, is expressed for each contact element as follows :

$$\text{gap } (E_c^{(ij)}) = (\mathbf{x}(\mathbf{z}^{(i)}) - \mathbf{x}(\mathbf{z}^{(j)}), \mathbf{N}(E_c^{(ij)}) \geq 0. \tag{5}$$

3.5 Constitutive laws for contact and friction

A constitutive law for normal contact is considered in order to link normal reactions to the gaps measured at contact elements. This constitutive law has the form of a penalty law, and is regularized by a quadratic part for very small penetrations in order to stabilize the contact algorithm (see Fig. 6). The penalty coefficient is adjusted for each proximity zone in order to limit the maximum penetration for each zone. As far as the friction is concerned, we use a Coulomb's law including a small reversible elastic displacement before sliding occurs.

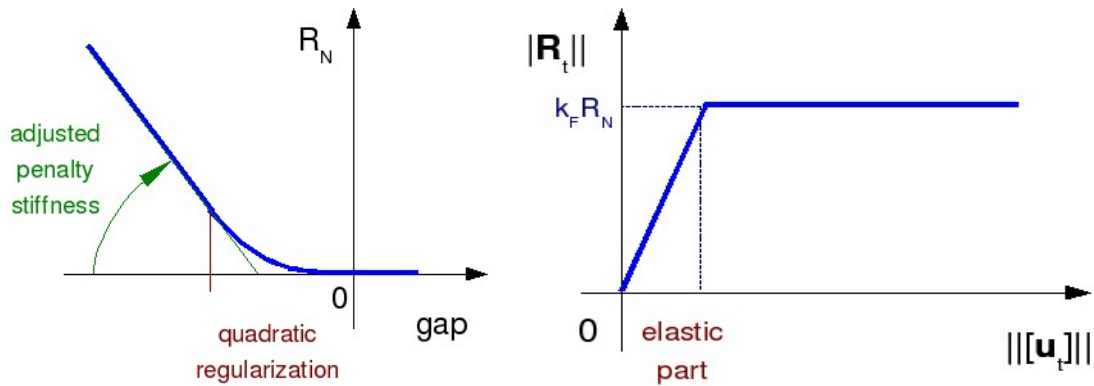

Figure 7. Normal contact law and tangential friction law used for interactions.

3.6 Global algorithm to solve the nonlinear problem

To solve the problem for each loading increment, imbricated loops are introduced to iterate on the different nonlinearities. For each increment, with a first level loop, iterations are made on the process of contact determination to generate contact elements. Then, these contact elements being fixed, iterations are made at a second level on the normal contact directions. The third level loop is dedicated to iterations of a Newton-Raphson algorithm on all other nonlinearities (contact status, friction direction, finite strains). Because of these three embedded loops, the total number of iterations may be high if a good rate of convergence is not obtained for the Newton-Raphson algorithm. Models characteristics need to be carefully adjusted for this purpose.

## 4. COMPUTATION OF THE INITIAL CONFIGURATION

4.1 Way of determining the initial configuration

One important task of the simulation is to determine the unknown initial geometry of the woven structure. Instead of simulating step by step the actual weaving process, which would require a large amount of CPU time, the initial configuration is computed by starting from an arbitrary flat configuration where crossing tows are interpenetrating each other (Fig. 8), and moving gradually fibers until tows are moved apart from each other at crossings. For each crossing, the weaving pattern provides a superimposition order which indicates which tow should be above the other. The goal of the transient stage of computation of the initial configuration is to make this superimposition order fulfilled by fibers from different tows at crossings. To achieve this, while standard normal contact directions (see 3.4.1) are considered between fibers of the same tow, between fibers belonging to different tows, we take as normal contact direction a vertical direction oriented according to the superimposition order defined at the crossing. By this means fibers of crossing tows are moved step by step upon each other. Once tows do not interpenetrate any longer, standard normal contact directions are considered for all contact elements to find the equilibrium configuration.

4.2  Computation of the initial configuration for a plain weave and a twill weave sample

In the following, all applications are made for the same initial arrangement of 12 glass fiber tows (6 tows in both fill and warp directions), described on Table 1. Different tows are employed in the fill and warp direction so as to produce unbalanced fabrics.

Table 1. Characteristics of the studied samples.

| **Characteristics of the studied samples** | |
| --- | --- |
| Number of fibers in fill tows | 44 |
| Number of fibers in warp tows | 22 |
| Total number of fibers | 408 |
| Number of nodes | 35 112 |
| Number of dofs | 316 008 |
| Number of contact elements | around 80 000 |

Two weaving patterns, corresponding respectively to a plain weave and a twill weave, are applied to this initial arrangement made of 408 fibers (Fig. 8). During the process of determination of the initial geometry, around 80 000 contact elements are generated.

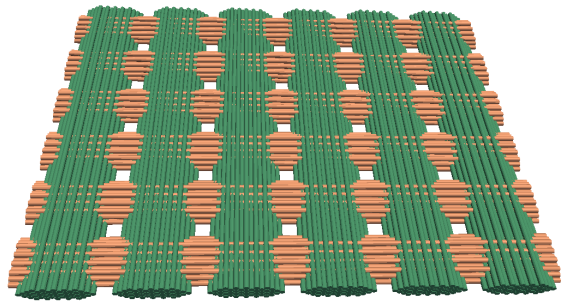

Figure 8. Initial configuration of tows before weaving.

The view of slices at some steps during the process of determination of the initial configuration (Fig. 9) shows how fibers of different tows are gradually moved apart.

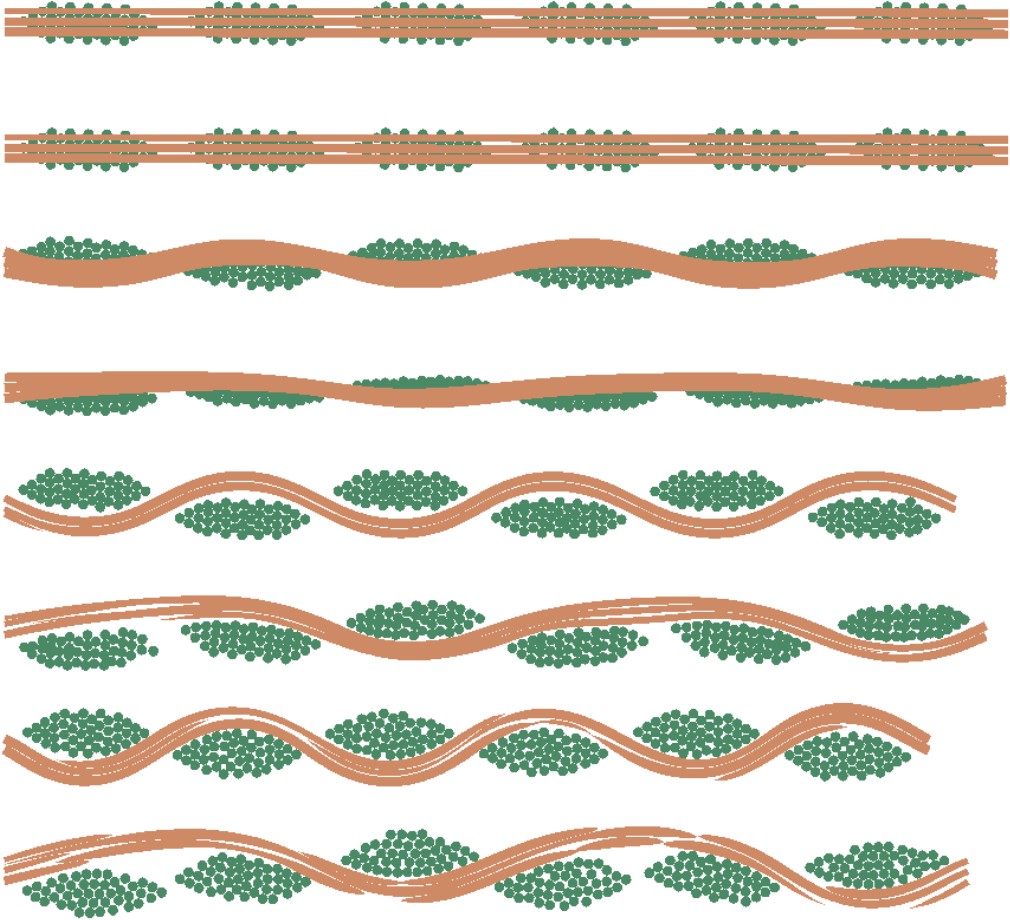

Figure 9. Slices at different steps during the process of determination of the initial configuration for the plain weave and the twill weave.

Many informations about the fabrics geometry are obtained as results of this initial simulation. Trajectories of both fibers and tows, varying shapes of tows cross-sections and local curvatures along fibers are very useful data which can be recovered from this first computation.

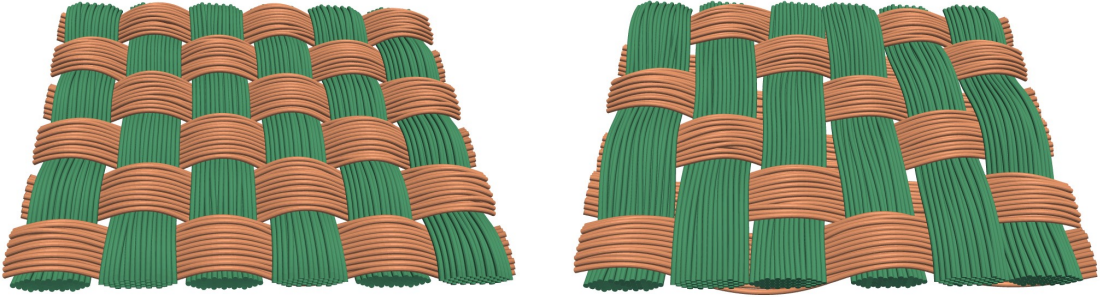

Figure 10. Computed initial configurations computed for plain and twill weaves.

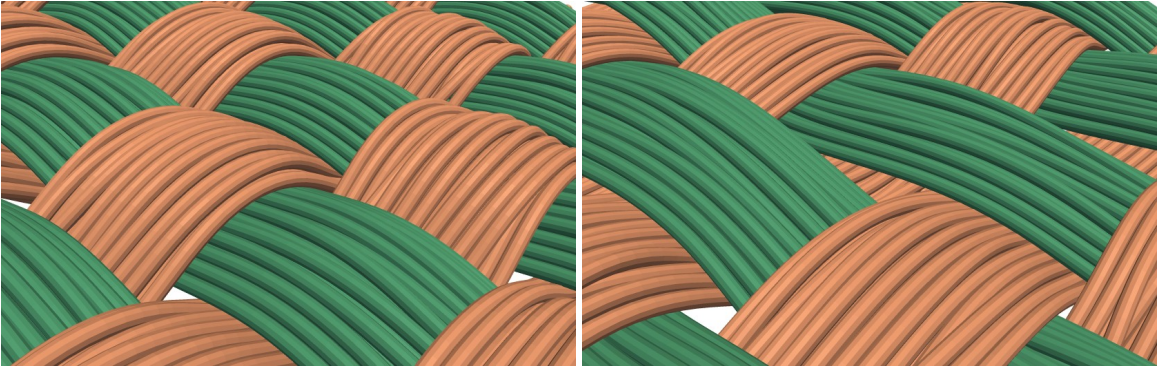

Figure 11. Details of the computed initial configurations for plain and twill weaves.

## 5. ADDITION OF AN ELASTIC MATRIX

In order to consider a composite sample, two layers of an elastic material are added on both sides of the fabric. The mesh of the layers is carried out automatically by the software. This mesh is relatively coarse compared to the size of the fibers, and overlaps slightly the external fibers of tows. Doing this way, the meshes of the fibers and of the matrix are non-conforming. To ensure the mechanical coupling between the fabric and the fibers, junction elements are generated in the overlapping region between the matrix and the fibers. These elements couple pairs of material particles, one belonging to a fiber and the other belonging to the matrix, by means of an elastic spring whose stiffness is calculated as function of the Young's modulus of the matrix.

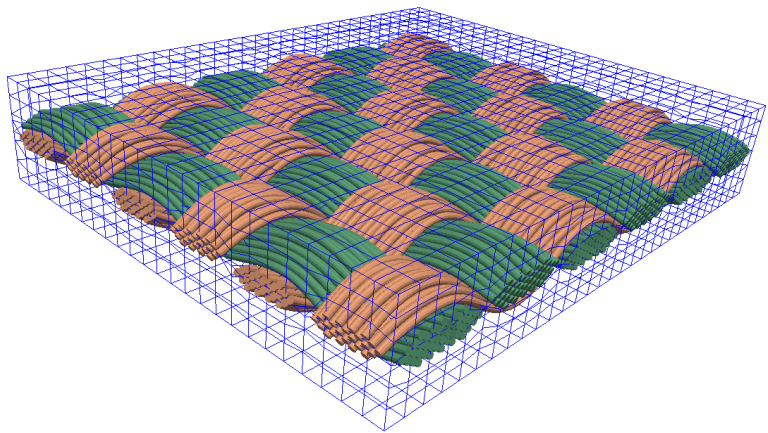

Figure 12. View of the solid mesh generated for the matrix on both sides of the fabric.

# 6. SIMULATION OF VARIOUS LOADING CASES

6.1 Presentation

Various loading cases can be simulated by applying different loads (displacements or forces) on the borders of the reconstructed samples of dry fabric or textile composite. Rigid bodies are introduced in order to drive globally sets of fiber ends or tow ends.

6.2 Biaxial tensile tests on dry fabric samples

Biaxial tensile tests are applied on the plain weave and on the twill weave samples. For these tests, two different ratios have been considered for elongations in warp and fill directions : a ratio of 1, with equal elongations in both directions, and a ratio of 0, with no elongation in warp direction. The corresponding loading curves are plotted on Fig. 13. They show the typical nonlinear behavior characterizing this kind of material.

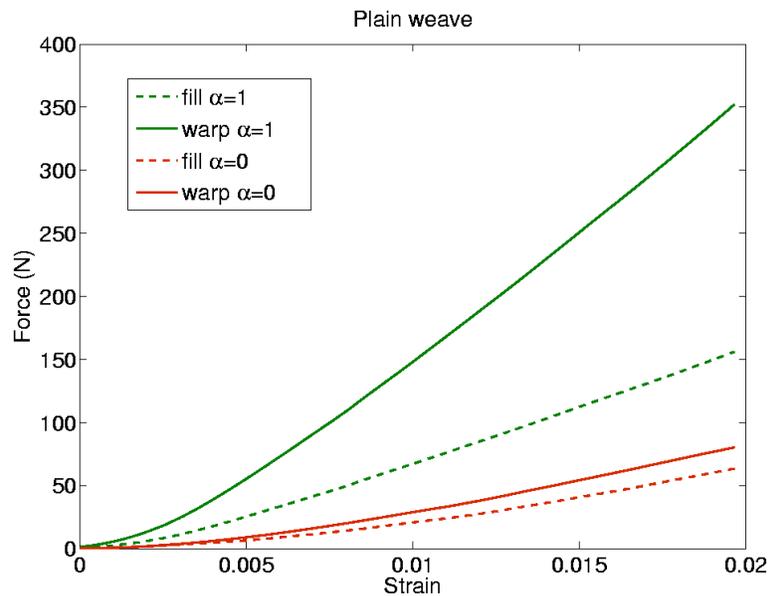

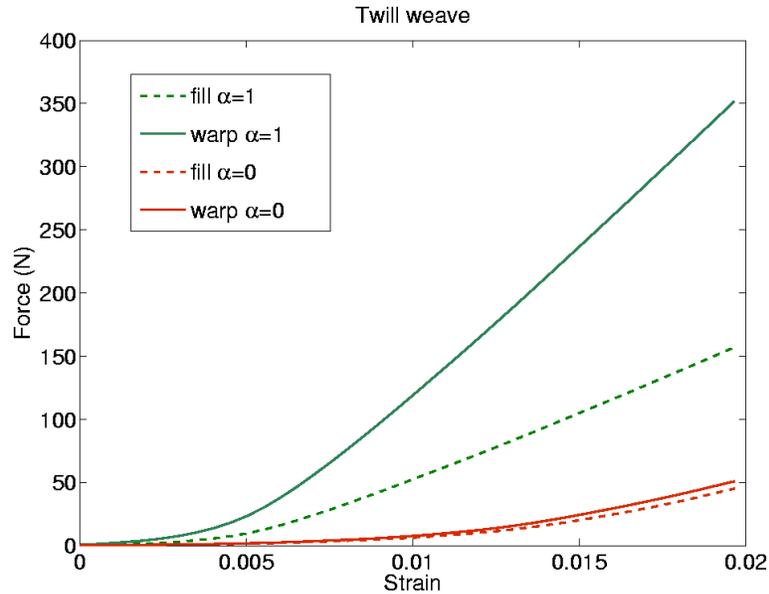

Figure 13. Loading curves in fill and warp directions for biaxial tensile tests on samples of unbalanced dry fabric for plain weave and twill weave.

The nonlinear effect at the start of the loading, which is stronger for the twill weave sample, can be mainly explained by a change in the fabric geometry due to a deformation of tows and a possible rearrangement of fibers in tows. This can be observed on Fig. 14, where slices of the twill weave at the beginning and at the end of the test show the changes in tows cross-section shapes and tows trajectories.

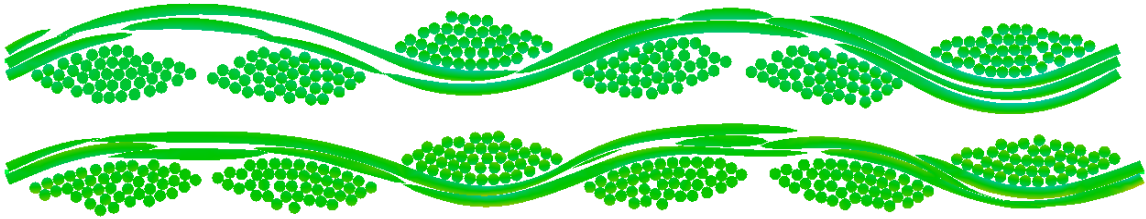

Figure 14. Slices of the twill weave fabric before (above) and after (below) a 2 % elongation test, showing deformation and rearrangement within tows cross-sections.

6.3  Shear test on the twill fabric composite

A shear test is performed on the composite textile sample for twill weave. The view of the deformed mesh is presented on Fig. 15. The shear deformation is possible until reaching a kind of locking when neighboring parallel tows come to contact. On a zoom view on Fig. 16, the deformation of tows can be observed.

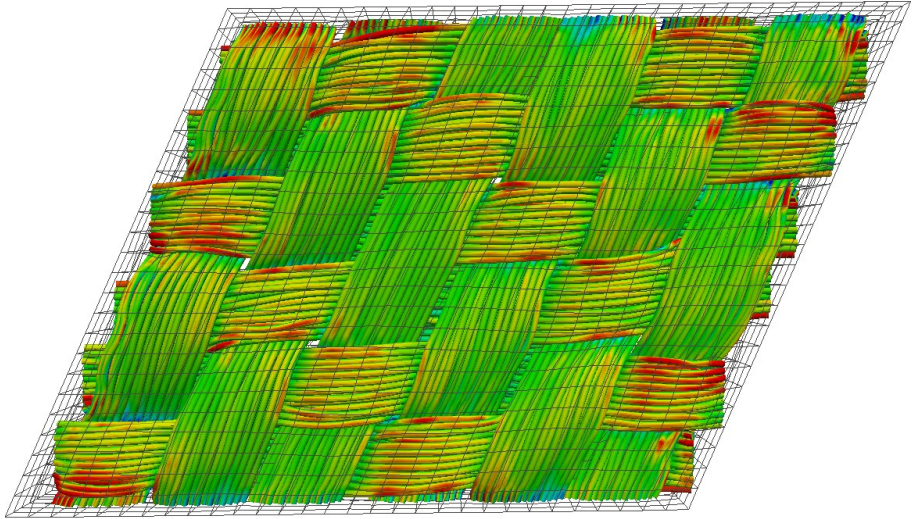

Figure 15. View of the textile composite sample submitted to a shear test.

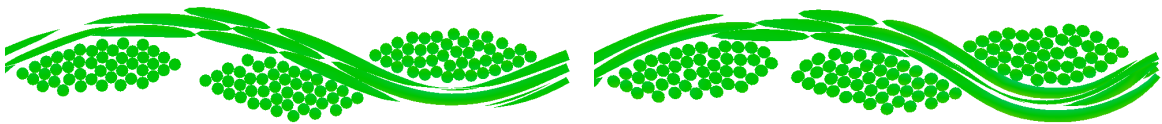

Figure 16. Slices at the beginning and at the end of the shear test showing changes is the shape of tows cross-sections.

6.4 Bending test on the twill fabric composite

The last test conducted on the composite sample is a bending test, for which opposite rotations are imposed to two borders of the sample. A global view of the sample is presented on Fig. 17, and a corresponding slice on Fig. 18. This test shows the ability of the model to handle large displacements.

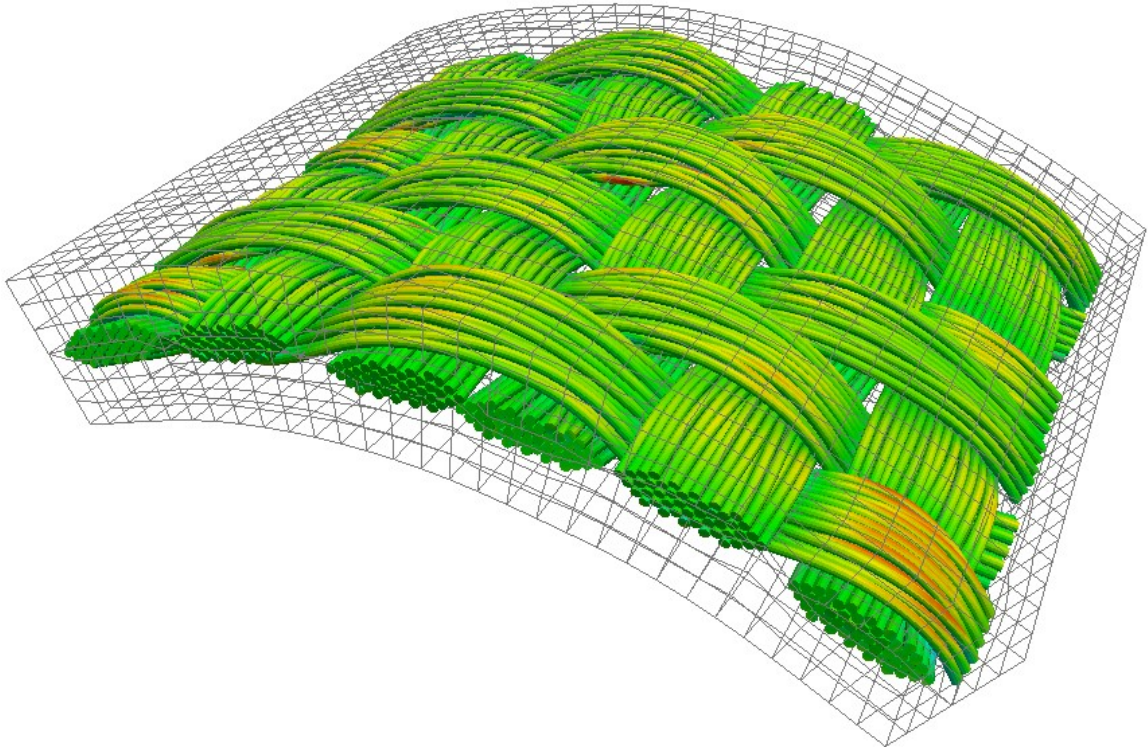

Figure 17. Global view of the composite sample after bending.

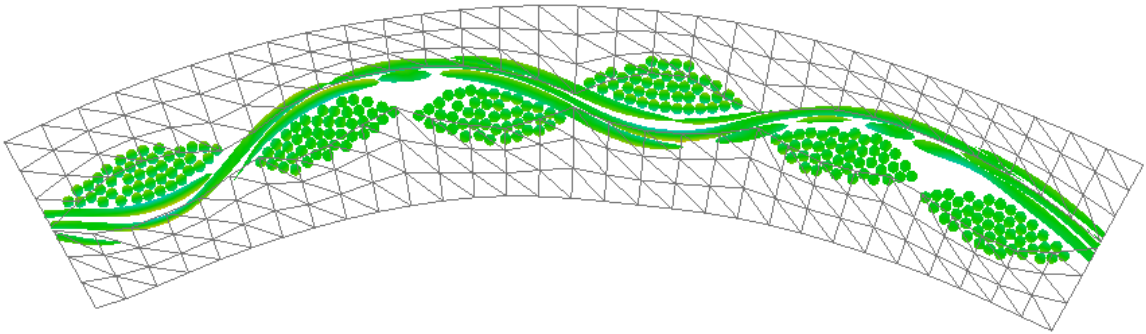

Figure 17. Slice of the composite sample submitted to bending.

## 7.  CONCLUSIONS

A general approach for the simulation by the finite element method of the mechanical behavior of fabrics and fabric-reinforced composites has been presented. This approach, formulated in a large displacements and finite strains framework, is based on the representation of all fibers constituting woven structures by means of 3D beam models, and on the taking into account of contact-friction interactions between fibers.

The detection of the numerous contacts occurring in general assemblies of fibers is one of the key points of the model. Based on the determination at a first level of proximity zones between fibers, and on the construction of intermediate geometries to approximate the actual contact zones, the method generates automatically contact elements made of pairs of material particles. This method is general enough to apply to many contact configurations as encountered between fibers in woven materials. Since the process of determination of contact elements depends on the relative positions of fibers, it has to be repeated during the solving of the problem for each loading step.

Robust models and efficient algorithms are required to get a good convergence rate for the solving of the nonlinear problem. Thanks to these optimized models and algorithms, samples of woven fabrics involving few hundreds of fibers and about 100 000 contact elements can be studied with a reasonable CPU time.

The model is first applied to the determination of the unknown initial geometry, by making the superimposition order at crossings defined by the weaving pattern be gradually fulfilled by fibers. Meaningful informations related to the yarns trajectories and shapes, and to the trajectories and curvatures of fibers, are obtained as results of this initial simulation.

In order to create a sample of textile composite, an elastic matrix is added to the numerically manufactured fabric, and is automatically meshed. Various loading cases can be than applied to the textile composite sample in order to identify its behavior under different kind of solicitations. Thanks to the nonlinearities considered in the model, and its ability to handle large loading increments, typical nonlinear effects of the complex behavior of textile materials can be rendered for a wide range of loadings.

The proposed simulation, based on the identification of very few parameters (mechanical properties of fibers, definition of tows arrangement and of the weaving pattern), appears as a suited tool to describe and to understand complex phenomena occurring at the scale of fibers, and to predict the complex global mechanical behavior of textile composite materials at a macroscopic scale, as well as very localized phenomena such as breakings at the microscopic scale of fibers.